\documentstyle[11pt,newpasp,twoside,epsf]{article}
\def\edcomment#1{\iffalse\marginpar{\raggedright\sl#1\/}\else\relax\fi}
\reversemarginpar

\begin{document}
\title{Deep Blazar Surveys}
\author{Paolo Padovani}
\affil{Space Telescope Science Institute, 3700 San Martin Drive, Baltimore,
MD, 21218, USA\\
Affiliated to the Astrophysics Division, Space Science Department, European
Space Agency\\
On leave from Dipartimento di Fisica, II Universit\`a di Roma ``Tor Vergata'',
Via della Ricerca Scientifica 1, I-00133 Roma, Italy}

\begin{abstract}
I address the need for deep blazar surveys by showing that our current
understanding of blazars is based on a relatively small number
of intrinsically luminous sources. I then review the on-going deeper surveys,
addressing in particular their limits and limitations. Finally, I present some
preliminary results on the evolutionary properties of faint blazars as derived
from the Deep X-ray Radio Blazar Survey (DXRBS).
\end{abstract}

\section{Introduction}

Blazars are the most extreme variety of Active Galactic Nuclei (AGN) known.
Their signal properties, discussed in detail in this volume, include
irregular, rapid variability; high optical polarization; core-dominant radio
morphology; apparent superluminal motion; flat ($\alpha_{\rm r} \la 0.5$)
radio spectra; and a broad continuum extending from the radio through the
gamma-rays (e.g., Urry \& Padovani 1995). Blazar properties are consistent
with relativistic beaming, that is bulk relativistic motion of the emitting
plasma at small angles to the line of sight (as originally proposed by
Blandford \& Rees in 1978), which gives rise to strong amplification and
collimation in the observer's frame. It then follows that an object's
appearance depends strongly on orientation. Hence the need for ``Unified
Schemes'', which look at intrinsic, isotropic properties, to unify
fundamentally identical (but apparently different) classes of AGN.

The blazar class includes flat-spectrum radio quasars (FSRQ) and BL Lacertae
objects. These are thought to be the ``beamed'' counterparts of high- and
low-luminosity radio galaxies, respectively. The main difference between the
two blazar classes lies in their emission lines, which are strong and
quasar-like for FSRQ and weak or in some cases outright absent in BL Lacs. The
current view is that there is actually a continuity of at least some
properties between the two classes, so the distinction between a BL Lac and an
FSRQ can be somewhat blurred (Landt et al., these proceedings).

Due to their peculiar orientation with respect to our line of sight, blazars
represent a very rare class of objects, making up considerably less than 5\%
of all AGN (Padovani 1997). As a consequence, all existing blazar samples
were, until very recently, relatively small and, due also to the difficulty in
identifying them, at high fluxes. It then follows that {\it our understanding
of the blazar phenomenon is mostly based on a relatively small number of
intrinsically luminous sources, which means we have only sampled the tip of
the iceberg of the blazar population.} For example, the radio luminosity
function (LF) of FSRQ derived by Urry \& Padovani (1995), although based on 52
sources (the best that could be done at the time), included only one source at
$L_{\rm r} < 10^{26.5}$ W Hz$^{-1}$, the power which coincides roughly with
the predicted flattening of the LF based on unified schemes (see
\S~7). Moreover, only in the limited range $10^{26.9} < L_{\rm r} < 10^{27.7}$
W Hz$^{-1}$ was the statistics good enough to have more than one source per
bin! The need for deeper, larger blazar samples is obvious.

\section{The ``Classical'' Blazar Samples}

Before I discuss the on-going, deeper blazar surveys, I summarize here the
basic facts about the ``classical'' blazar samples, the ones we all know and
love and on which our knowledge of blazars is based.
\medskip

\leftline{\sl BL Lacs}

\begin{itemize}

\item 1 Jy, radio flux-limited, $f_{\rm 5 GHz} > 1$ Jy, with radio
spectral index cut $\alpha_{\rm r} \le 0.5$, $V < 20$; complete sample includes 34
objects (Stickel et al. 1991);

\item EMSS, X-ray flux-limited, $f_{\rm 0.3-3.5 keV} \ga 2 \times 10^{-13}$
erg/cm$^{2}$/s; complete sample includes 41 objects (Stocke et al. 1991;
Rector et al. 2000);

\item IPC Slew, X-ray flux-limited, $f_{\rm 0.3-3.5 keV} \ga 2 \times 10^{-12}$
erg/cm$^{2}$/s; complete sample includes 51 objects (Perlman et al. 1996).

\end{itemize}
 
\medskip
\leftline{\sl Flat-spectrum Radio Quasars}

\begin{itemize}

\item 2 Jy, radio flux-limited, $f_{\rm 2.7 GHz} > 2$ Jy; complete sample
includes 52 objects (Wall \& Peacock 1985; di Serego Alighieri et al. 1994).

\end{itemize}

\section{New Blazar Samples}

Many groups are tackling the problem of assembling deeper, sizable blazar
samples, for the reasons discussed above: number statistics and limiting
fluxes. Most of these samples take advantage of the fact that blazars are
relatively strong radio and X-ray sources and use a double radio/X-ray
selection method (unlike the ``classical'' samples). Another difference lies
in the identification process. When dealing with catalogs of up to $\sim
1,000$ sources, one can obtain an optical spectrum of all of them and identify
the blazars. With the deeper, larger catalogs available today, with numbers
$\ga 100,000$ and reaching the millions, this becomes impossible without
access to dedicated facilities or unlimited resources. Hence the need to
increase the efficiency (e.g., via cross-correlation methods) to restrict the
number of blazar candidates down to a manageable number.

Since not all groups could present their results at this conference, I
summarize here the main on-going surveys, in chronological order. I make no
claim of completeness, but I have tried my best to include the largest,
deepest samples.

\medskip
\medskip
\leftline{\sl BL Lacs}

\begin{itemize}

\item DXRBS (Deep X-ray Radio Blazar Survey); uses radio (GB6, PMN)/X-ray (WGA
[ROSAT PSPC]) selection; survey limits are $f_{\rm 5 GHz} \ga 50$ mJy, $f_{\rm
0.1-2.4 keV} \ga 2 \times 10^{-14}$ erg/cm$^{2}$/s, with a cut in radio
spectral index $\alpha_{\rm r} \le 0.7$; complete sample includes 37 objects
(43 in whole sample) and is $\sim 90\%$ identified as of October 2000 (Perlman
et al. 1998; Landt et al. 2001; and papers in preparation); 

\item RGB (ROSAT All Sky Survey [RASS]-Green Bank) sample; uses radio
(GB6)/X-ray (RASS) selection, with an optical limit; survey limits are $f_{\rm
5 GHz} \ga 20$ mJy, $f_{\rm 0.1-2.4 keV} \ga 3 \times 10^{-13}$
erg/cm$^{2}$/s, $B<18$; complete sample includes 33 objects (127 in whole
sample) and is $\sim 94\%$ identified (Laurent-Muehleisen et al. 1998; 1999);

\item REX (Radio-Emitting X-ray) sample; uses radio (NVSS)/X-ray (ROSAT PSPC)
selection; survey limits are $f_{\rm 1.4 GHz} > 5$ mJy, $f_{\rm 0.1-2.4 keV}
\ga 3 \times 10^{-14}$ erg/cm$^{2}$/s; sample includes 72 objects, $\sim 30\%$
identified; subsample of $\sim 40$ objects with $f_{\rm 0.1-2.4 keV} \ga 4
\times 10^{-13}$ erg/cm$^{2}$/s is $\sim 90\%$ identified (Caccianiga et
al. 1999; 2000; Caccianiga et al., these proceedings);

\item ``Sedentary'' Survey; uses radio (NVSS)/X-ray (RASS)/optical (APM,
COSMOS) selection; survey limits are $f_{\rm 1.4 GHz} > 3.5$ mJy, $f_{\rm
0.1-2.4 keV} \ga 10^{-12}$ erg/cm$^{2}$/s; two-point spectral index selection
as well, $\alpha_{\rm rx} \la 0.56$, $\alpha_{\rm ro} > 0.2$, to select a
region populated by high-energy peaked BL Lacs (HBL) at $\sim 85\%$
level. Sample includes 155 candidates, $\sim 70\%$ identified, but high
efficiency expected (Giommi, Menna \& Padovani 1999; Giommi et al., these
proceedings);

\item FIRST Flat Spectrum sample; uses radio (FIRST, GB6) selection, with an
optical limit; survey limits are $f_{\rm 1.4 GHz} > 35$ mJy, $f_{\rm 5
GHz} > 20$ mJy, $B < 19$, with a cut in radio spectral index $\alpha_{\rm r} <
0.5$. Sample includes 87 sources and is $\sim 84\%$ identified
(Laurent-Muehleisen et al., in preparation).
\end{itemize}

\medskip

\leftline{\sl Flat-spectrum Radio Quasars}
 
\begin{itemize}
\item Parkes 0.25 Jy sample; uses radio selection (PKS); survey limit is
$f_{\rm 2.7 GHz} > 250$ mJy, with a cut in radio spectral index $\alpha_{\rm
r} \le 0.4$; 444 sources, sample is $100\%$ identified, in the process of
being published (Shaver et al. 1996; Hook et al. 1999; Jackson \& Wall, these
proceedings);

\item DXRBS (Deep X-ray Radio Blazar Survey); uses radio (GB6, PMN)/X-ray (WGA
[ROSAT]) selection; survey limits are $f_{\rm 5 GHz} \ga 50$ mJy, $f_{\rm
0.1-2.4 keV} \ga 2 \times 10^{-14}$ erg/cm$^{2}$/s, with a cut in radio
spectral index $\alpha_{\rm r} \le 0.7$; complete sample includes 187 objects
(193 in whole sample) and is $\sim 90\%$ identified as of October 2000 (Perlman
et al. 1998; Landt et al. 2001; and papers in preparation);

\item FIRST Flat Spectrum sample; uses radio (FIRST, GB6) selection, with an
optical limit; survey limits are $f_{\rm 1.4 GHz} > 35$ mJy, $f_{\rm 5
GHz} > 20$ mJy, $B < 19$, with a cut in radio spectral index $\alpha_{\rm r} <
0.5$; 332 sources, sample is $\sim 84\%$ identified (Laurent-Muehleisen et
al., in preparation).
\end{itemize}

\section{Parameter Space Coverage}

It is important to assess what regions of parameter space these various
surveys are sensitive to, in order to understand what constraints they can or
cannot put on blazar demographics. Given the double (radio/X-ray) selection
criteria of most of the new surveys and the fact that the ``classical'' blazar
samples were either radio or X-ray selected, I analyze how
these samples cover the radio--X-ray flux plane.

\begin{figure}
\epsfxsize=9.0cm 
\hspace{1.5cm}\epsfbox{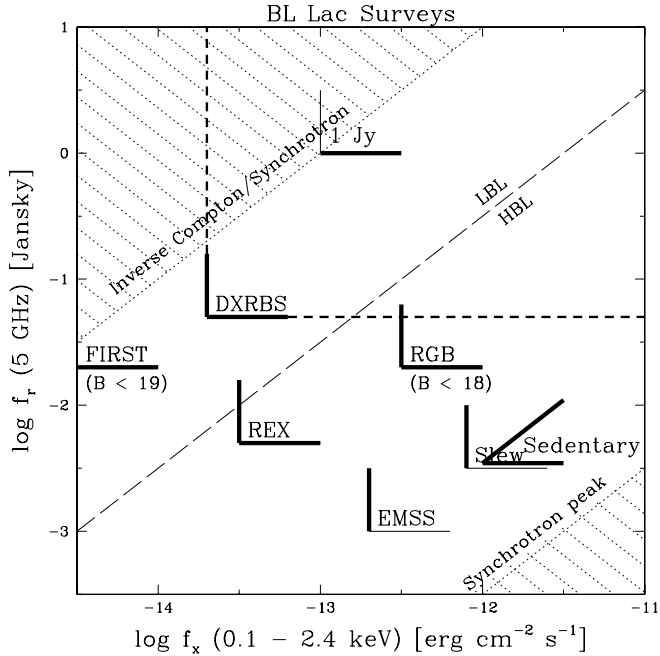} 
\caption[h]{The sampling of the radio flux -- X-ray flux plane by different BL
Lac surveys. Thick lines represent ``hard'' survey limits, while thin lines
are the fluxes reached in a band other than the one of selection. Sources
belonging to a given survey occupy a region of the plane whose bottom-left
corner is indicated by the thick/thin lines, as exemplified for DXRBS
(short-dashed lines). The long-dashed line divides HBL from LBL, while the
hatched regions represent the ``forbidden'' zones, where no known BL Lacs have
been found so far. See text for more details.}
\end{figure}



This is shown in Figure 1 for BL Lacs. Every survey is characterized by one
(or two) flux limits (thick lines), while the smallest flux reached by a
sample in a band other than the one of selection is given by a thin line. For
example, the EMSS BL Lacs reach $f_{\rm x} \sim 2 \times 10^{-13}$
erg/cm$^{2}$/s (thick line), by default the X-ray selection limit (actually,
the faintest of various limits, due to the nature of the survey). A limit in
one band translates into a limit in the other and in this case the radio
faintest EMSS BL Lac has a flux $f_{\rm r} \sim 1$ mJy (thin line). The
sources of a given survey occupy a region of the flux-flux plane whose
bottom-left corner is shown in the figure. 

The long-dashed line in the figure (X-ray-to-radio flux ratio $f_{\rm
x}/f_{\rm r} = 10^{-11.5}$ erg/cm$^{2}$/s/Jy or $\alpha_{\rm rx} \sim 0.78$)
divides HBL from low-energy peaked BL Lacs (LBL). Although this distinction
might sound arbitrary, there is convincing evidence that HBL are
synchrotron-dominated in the X-ray band, unlike LBL where two components (or
only one, inverse Compton emission) might coexist (Padovani \& Giommi 1996).
The parallel dotted lines (lines of constant $f_{\rm x}/f_{\rm r}$) represent
the known range in $f_{\rm x}/f_{\rm r}$ for BL Lacs, which I derived from
available X-ray and radio data. This is $10^{-13} \la f_{\rm x}/f_{\rm r} \la
10^{-8.5}$ erg/cm$^{2}$/s/Jy (or $0.4 \la \alpha_{\rm rx} \la 1$). No known BL
Lacs occupy the hatched regions. I believe that this is not mainly a selection
effect, but that there are physical reasons for this. The limit at the low end
of the $f_{\rm x}/f_{\rm r}$ range (marked ``Inverse Compton/Synchrotron'' in
the figure) is likely due to the fact that in extreme LBL sources the X-ray
band is dominated by inverse (synchrotron self-) Compton emission, the radio
emission is synchrotron, and the ratio of the two is proportional to the ratio
of photon density, $W$, to $B^2$, where $B$ is the magnetic field strength.
There are probably physical reasons why $W/B^2$ cannot reach indefinitely low
values in blazars (although I cannot exclude that sources with smaller $f_{\rm
x}/f_{\rm r}$ exist). At the other end, the higher $f_{\rm x}/f_{\rm r}$, the
larger the peak frequency of the synchrotron emission, $\nu_{\rm peak}$, in
extreme HBL sources. And even in this case there are plausible physical
reasons that limit $\nu_{\rm peak}$, which depends on the maximum electron
energy (Ghisellini 1999).

A few interesting points can be made about the position of the various surveys
on the $f_{\rm r} - f_{\rm x}$ plane. First, it is clear why we came to think
of radio-selected (RBL) and X-ray selected (XBL) BL Lacs as different types of
sources: the 1 Jy and EMSS surveys sample vastly different regions of
parameter space. Based on these two ``classical'' surveys it was hard to see
that there was a distribution of synchrotron peak frequencies of which the 1
Jy and EMSS samples represented the two extremes. With the Slew sample we
started to bridge the gap, as a few Slew BL Lacs are LBL and ``intermediate'',
but it was not until the more recent surveys (DXRBS, REX, RGB), whose limits
straddle the HBL/LBL division, that we realized that intermediate BL Lacs
indeed existed in sizable numbers.
Second, it is important to realize the limitations of surveys with
double (radio/X-ray) flux limits. A survey whose limits fall quite far from
the two dotted lines will not provide a complete picture of the BL Lac
population. For example, the REX survey cannot provide BL Lac radio number
counts to be compared with the predictions of a beaming model based on the 1
Jy sample, simply because it does not include all the BL Lacs above its radio
limit (as it misses all those above the radio limit but below the X-ray
limit). For the complementary reason, neither can REX provide X-ray number
counts to be compared with the predictions from a beaming model based on the
EMSS sample. REX will provide radio number counts for HBL, given its proximity
to the HBL/LBL dividing line, and X-ray number counts for LBL (as it detects
all LBL above its X-ray flux limit). The same arguments apply to the RGB
survey, which has the further problem of an optical limit ($B < 18$). This
implies that only BL Lacs with radio-optical spectral index $\alpha_{\rm ro} <
\alpha_{\rm ro}({\rm lim})$, where $\alpha_{\rm ro}({\rm lim})$ depends on
radio flux (and is $\sim 0.4$ at the survey limit, for example) will be
included. In the case of DXRBS, on the other hand, being relatively close to
the leftmost boundary of the BL Lac region, the X-ray flux limit is not as
important and can therefore be considered ``almost'' radio flux-limited
only. The ideal sample, of course, has only one, faint, flux limit. FIRST does
not have any X-ray cut but unfortunately the optical limit ($B< 19$) implies,
as for RGB, that only BL Lacs with radio-optical spectral index $\alpha_{\rm
ro}$ flatter than a given value (which depends on radio flux) will be
included.

\begin{figure}
\epsfysize=9.0cm 
\hspace{1.5cm}\epsfbox{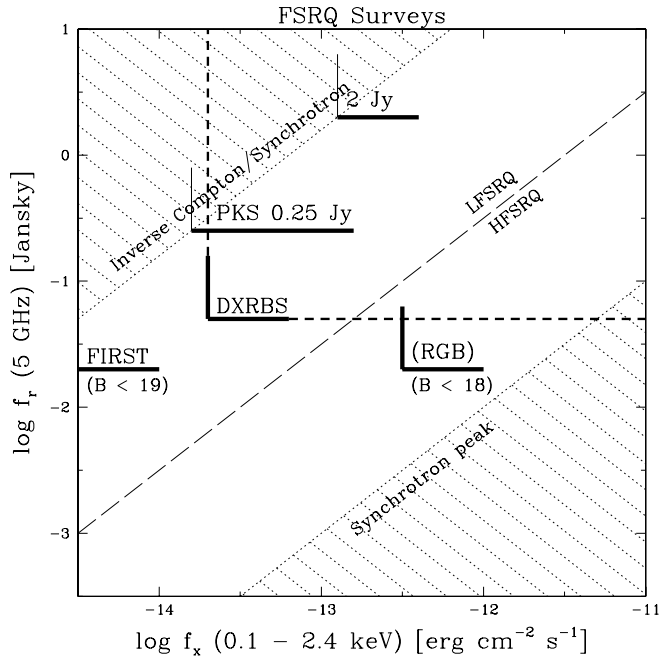} 
\caption[h]{The sampling of the radio flux - X-ray flux plane by different
FSRQ surveys. Thick lines represent ``hard'' survey limits, while thin lines
are the fluxes reached in a band other than the one of selection. Sources
belonging to a given survey occupy a region of the plane whose bottom-left
corner is indicated by the thick/thin lines, as exemplified for DXRBS
(short-dashed lines). The long-dashed line divides HFSRQ (high-energy peaked
FSRQ) from LFSRQ (low-energy peaked FSRQ), while the hatched regions represent
the ``forbidden'' zones, where no known FSRQ have been found so far. See text
for more details.}
\end{figure}

The coverage of the radio--X-ray flux plane for FSRQ surveys is shown in
Fig. 2. As in Fig. 1, the parallel dotted lines represent the known range in
$f_{\rm x}/f_{\rm r}$. For FSRQ I find $10^{-13.2} \la f_{\rm x}/f_{\rm r} \la
10^{-10}$ erg/cm$^{2}$/s/Jy (or $0.6 \la \alpha_{\rm rx} \la 1.05$). The
long-dashed line divides high-energy peaked and low-energy peaked FSRQ (HFSRQ
and LFSRQ respectively) at $f_{\rm x}/f_{\rm r} = 10^{-11.5}$
erg/cm$^{2}$/s/Jy (or $\alpha_{\rm rx} \sim 0.78$). The existence of HFSRQ,
flat-spectrum quasars with synchrotron peak in the UV/X-ray band, was not
suspected until the first results of DXRBS (Perlman et al. 1998; Landt et
al. 2001; Perlman et al., these proceedings; Padovani et al., in
preparation). As shown in Fig. 2, in fact, only by reaching relatively faint
radio fluxes and by having X-ray information one can sample the HFSRQ region
of the plane. The 2 Jy sample had too high of a radio flux limit to include a
sizable number of FSRQ above the LFSRQ/HFSRQ line (only two sources in the
sample, in fact, have $\alpha_{\rm rx} < 0.78$, both of them with $f_{\rm x}
\sim 10^{-11}$ erg/cm$^{2}$/s). Note, however, that the region of the plane
occupied by HFSRQ is {\it smaller} than that occupied by HBL (compare the
position of the rightmost dotted line labeled ``Synchrotron peak'' in Fig. 1
and 2). For reasons we still do not understand, there seem to be no FSRQ with
synchrotron peak at energies as high as those reached by HBL (Padovani et al.,
in preparation).

Turning to the surveys themselves, I first note that RGB does not include
information on the radio spectral index, which is why it is in parentheses in
the figure. Padovani et al. (in preparation) have cross-correlated the RGB
sample with the NVSS to obtain radio spectral indices and extract the
FSRQ. The limitations of the RGB sample described above (due to its position
in the plane) still apply but given its radio/X-ray flux limits
this is the survey which is most suited to find HFSRQ. As discussed above
DXRBS, by being close to the leftmost dotted line, can be considered
``almost'' radio flux-limited only and therefore provides a sample of FSRQ
which can be used to test the predictions of unified schemes at low radio
fluxes. FIRST reaches even deeper fluxes but the optical limit ($B < 19$)
implies that it will not give a complete picture of the FSRQ population.

In summary, it is vital to understand what the various surveys can and cannot
provide and have their limits and limitations clear. In particular, {\it
surveys with more than one flux limit can provide a complete picture of the
blazar population only if the additional limits are relatively close to one
edge of the region of parameter space occupied by blazars.} I now turn to
analyze the preliminary results of DXRBS, the survey which I am directly
involved with and for which I have direct access to the data, in terms of
blazar demographics. 

\section{The Evolutionary Properties of DXRBS Blazars} 

The basic idea behind the Deep X-ray Radio Blazar Survey (DXRBS) is quite
simple: blazars are relatively strong X-ray and radio emitters so selecting
X-ray and radio sources with flat radio spectrum (one of their defining
properties) should be a very efficient way to find these rare sources. By
adopting a spectral index cut $\alpha_{\rm r} \le 0.7$ DXRBS: 1. selects all
FSRQ (defined by $\alpha_{\rm r} \le 0.5$); 2. selects basically 100\% of BL
Lacs; 3. excludes the large majority of radio galaxies. 

The survey limits are given in \S~3, while details on the selection technique
and identification procedures can be found in Perlman et al. (1998) and Landt
et al. (2001). Here I will just note that DXRBS is currently the faintest and
largest flat-spectrum radio sample with nearly complete ($\sim 90\%$ as of
October 2000) identification. Redshift information is available for $\sim
95\%$ of the identified sources.

The simplest way to study the evolutionary properties of a sample is through
the $V/V_{\rm max}$ test or, since the X-ray flux limit is a function of the
area, the $V_{\rm e}/V_{\rm a}$ test (Avni \& Bahcall 1980). Values of $V_{\rm
e}/V_{\rm a}$ significantly different from 0.5 indicate evolution, which will
be positive (i.e., sources were more luminous and/or more numerous in the
past) for values $>0.5$, or negative (i.e., sources were less luminous and/or
less numerous in the past) for values $<0.5$. Moreover, one can fit an
evolutionary model to the sample by finding the evolutionary parameter which
makes $V_{\rm e}/V_{\rm a} = 0.5$.

The DXRBS sky coverage (the area of sky surveyed as a function of X-ray flux)
has been derived by Paolo Giommi and Matteo Perri and will be used to derive
the evolutionary properties of the sample. I present here some preliminary
results based on the sample as of July 2000 ($\sim 30$ more sources have been
identified in August 2000 but are not included in this analysis). The sky
coverage is difficult to determine in the regions of the ROSAT PSPC field of
view affected by the rib structure ($13^{\prime} < {\rm offset} <
24^{\prime}$) That area, and the sources within, have therefore been excluded
from this analysis. Moreover, only sources with $f_{\rm r} > 51$ mJy have been
included since we still have not computed the sky coverage of the PMN survey
below this flux (this excludes however only a handful of objects). Table 1
gives the sub-sample, the mean $V_{\rm e}/V_{\rm a}$ value, $\langle V_{\rm
e}/V_{\rm a} \rangle$, the number of objects, and the best fit parameter
$\tau$ assuming a pure luminosity evolution of the type $P(z) = P(0)
exp[T(z)/\tau]$ (where $T(z)$ is the look-back time). The values $H_0 = 50$
km/s/Mpc and $q_0=0$ have been adopted. 

\begin{table}
\caption{DXRBS Evolutionary Properties}
\begin{center}
\begin{tabular}[h]{|l r r r |}
\hline
Sample & $\langle V_{\rm e}/V_{\rm a} \rangle$ ~~& N & $\tau$ ~~~~\\ 
\hline
All FSRQ & $0.58\pm0.03$  & 119 & $0.35^{+0.14}_{-0.08}$ \\
HFSRQ    & $0.71\pm0.05$  &  32 & $0.17^{+0.03}_{-0.02}$ \\
BL Lacs  & $0.57\pm0.05$  &  30 & \\
HBL      & $0.65\pm0.09$  &  11 & \\
LBL      & $0.52\pm0.07$  &  19 & \\
Unclassified ($z=1.5$) & $0.80\pm0.05$  &  39 & \\
\hline
\end{tabular}
\end{center}
\label{id}
\end{table}

The main results are the following: 

1. DXRBS FSRQ evolve; however, their
$\langle V_{\rm e}/V_{\rm a} \rangle$ value and evolutionary parameter $\tau$
reflect the fact that the sample is not completely identified (incompleteness
decreases $\langle V_{\rm e}/V_{\rm a} \rangle$). By restricting the analysis
to the HFSRQ (defined here by $\alpha_{\rm rx} \le 0.78$), a basically
complete sub-sample as most unidentified sources have $\alpha_{\rm rx} > 0.78$
(pending the effect of the k-correction on their $\alpha_{\rm rx}$ values)
$\langle V_{\rm e}/V_{\rm a} \rangle$ increases and the value of $\tau$
becomes consistent (within $\sim 2 \sigma$) with that of 2 Jy FSRQ (Urry \&
Padovani 1995). 

2. DXRBS BL Lacs do not evolve, i.e., their $\langle V_{\rm
e}/V_{\rm a} \rangle$ value is not significantly different from 0.5 (and
consequently $\tau \ga 1$). (The results for BL Lacs are however more
uncertain because of the smaller number statistics and the fact that $\sim
30\%$ of them have no redshift; $z=0.4$ was assumed in this
case). 

3. the $\langle V_{\rm e}/V_{\rm a} \rangle$ values for HBL and LBL are
not significantly different. This is a new result, which contradicts the
commonly accepted fact that HBL and LBL have different evolutionary
properties. Notice that {\it for the first time} we can study the evolution of
HBL and LBL {\it within the same sample}. Previous comparisons had been made
between the 1 Jy (radio-selected) and the EMSS samples
(X-ray-selected). Admittedly, the errors on the $\langle V_{\rm e}/V_{\rm a}
\rangle$ values are rather large but since, as noticed above, the still
unidentified sources are mostly of the LBL type, completion of the
identification process will likely {\it decrease} the difference between the
HBL and LBL values. 

4. $\langle V_{\rm e}/V_{\rm a} \rangle$ for the still
unclassified sources which, based on our results so far, will be for the most
part FSRQ, is quite high (assuming $z=1.5$, the mean value for the FSRQ); this
implies that when these sources will be identified and included in the whole
FSRQ sample, the FSRQ $\langle V_{\rm e}/V_{\rm a} \rangle$ will likely reach
that of the HFSRQ.

The $V_{\rm e}/V_{\rm a}$ test is a simple way to study the evolutionary
properties of a sample. To move to the demographics one needs number counts
or, if complete redshift information is available, the luminosity function. I
will address these in turn for the DXRBS BL Lacs and FSRQ.

\section{DXRBS BL Lac Number Counts}
\begin{figure}
\epsfysize=9.0cm 
\hspace{1.5cm}\epsfbox{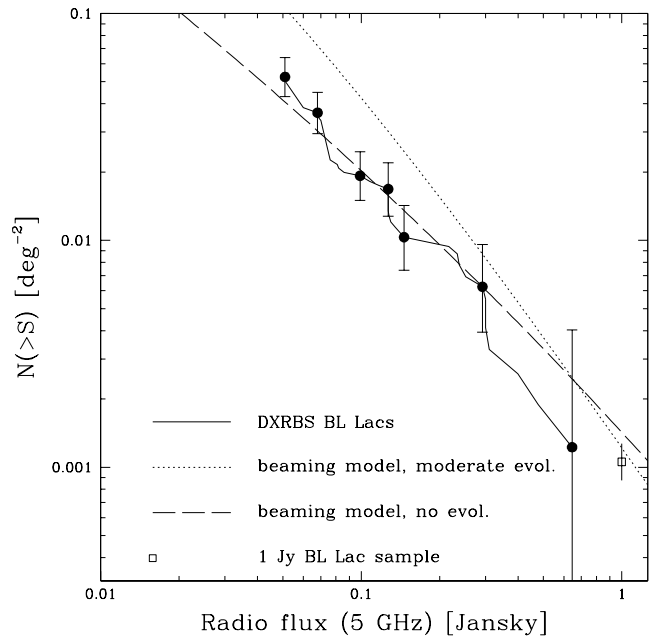} 
\caption[h]{The (preliminary) radio integral number counts at 5 GHz of DXRBS
BL Lacs (solid line and filled points) compared to the predictions of a
beaming model with the moderate evolution of the 1 Jy BL Lacs (dotted line)
and no evolution (dashed line). The empty square represents the surface
density of the 1 Jy BL Lacs. Error bars correspond to $1\sigma$ Poisson errors
and are shown only for a few selected points for clarity. The drop in the
observed counts at high fluxes is due to the serendipitous nature of DXRBS.
All the ROSAT targets, in fact, have been excluded and these were mostly
well-known, high-flux sources.} 
\end{figure}
For the past 10 years or so the only sizable, complete, radio-selected sample
of BL Lacs has been the 1 Jy sample (Stickel et al. 1991). The predictions of
relativistic beaming have been tested and tuned to this sample and constraints
on beaming parameters (Lorentz factor distribution, angles) have been
derived. We can now test unified schemes on a sample which reaches $\sim 20$
times fainter radio fluxes. Given the fact that redshifts are still missing
for $\sim 30\%$ of the DXRBS BL Lacs we start by deriving the radio number
counts.

Figure 3 shows the (preliminary) integral number counts at 5 GHz for the DXRBS
BL Lacs down to $\sim 50$ mJy, compared to the predictions of unified schemes
based on a fit to the 1 Jy LF (Urry \& Padovani 1995). The DXRBS counts have
been corrected for incompleteness by scaling them up by $15\%$. The dotted
line assumes the best-fit 1 Jy evolution ($\tau=0.32^{+0.27}_{-0.08}$). Note
that the $V/V_{\rm m}$ value for the 1 Jy BL Lacs is $0.60\pm0.05$, i.e., a
departure from the non evolutionary case significant only at the $2\sigma$
level. For this reason, and because the $V_{\rm e}/V_{\rm a}$ results for
DXRBS are at present consistent with no evolution, I also show the surface
density of BL Lacs predicted assuming no evolution. Fig. 3 shows that the
preliminary number counts agree with the no evolution case, in agreement with
the $V_{\rm e}/V_{\rm a}$ results. There are a couple of caveats, however,
which should be kept in mind. First, the identification process of the DXRBS
sample is not complete yet. This has been taken into account by scaling the
counts up appropriately but most of the unidentified sources are the faint end
so the shape of the counts could change. Second, the definition of a BL Lac
for the 1 Jy and DXRBS samples is different, the latter being less restrictive
following March\~a et al. (1996). This implies that the comparison between
predictions and observations should be restricted to the DXRBS BL Lacs which
fulfill the 1 Jy definition. As these make up $\sim 70\%$ of the sample,
however, this should not make much of a difference.

\section{DXRBS FSRQ Luminosity Function}

The situation for FSRQ is better, both because of the better statistics and
the fact that redshifts are available for all objects. In this case we can
then derive directly the luminosity function and compare it with what expected
from unified schemes. I take into account the fact that the identification is
not complete yet by applying the best-fit evolution derived from the complete
subsample of HFSRQ to the whole sample. Keeping this in mind, Figure 4
presents the (preliminary) local radio luminosity function (de-evolved to zero
redshift using the best-fit evolution) for the DXRBS FSRQ. The predictions of
unified schemes based on a fit to the 2 Jy LF (Urry \& Padovani 1995) are also
shown (solid line). A few interesting points can be made: 1. the 2 Jy and
DXRBS LFs are in good agreement in the region of overlap; 2. DXRBS has much
better statistics: the two lowest bins of the 2 Jy LF contain only one object
each, while the number of DXRBS sources in the same bins is $\sim 20 - 30$;
3. the DXRBS LF reaches powers more than one order of magnitude smaller than
those reached by the 2 Jy LF, as expected given the much fainter ($\sim 30$)
flux limit; 4. the DXRBS LF is in (amazingly!) good agreement with the
predictions of unified schemes; 5. we are getting close to the limits of the
FSRQ ``Universe''; as FSRQ are thought to be the beamed counterparts of
high-power radio galaxies, their luminosity function should end at relatively
high powers. Assuming that the value inferred from the fit to the 2 Jy LF is
correct (solid line in the figure, based on the 2 Jy LF of Fanaroff-Riley type
II radio galaxies; see Urry \& Padovani 1995), then DXRBS is approaching that
value. 

\begin{figure}
\epsfysize=7.9cm 
\hspace{1.5cm}\epsfbox{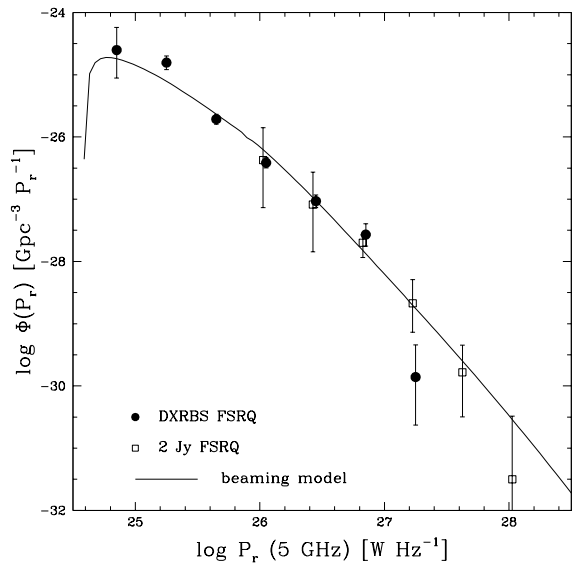} 
\caption[h]{The (preliminary) radio luminosity function of DXRBS FSRQ (filled
points) compared to the predictions of a beaming model based on the 2 Jy
luminosity function and evolution (solid line). The open squares represent the
2 Jy luminosity function. Error bars correspond to $1\sigma$ Poisson errors.}
\end{figure}

\section{Even Deeper Surveys?}

What is in store for the future? Will we be able to go even deeper in our
quest for blazars, to probe the even less powerful sources? It will not be
easy. Consider in fact a radio survey reaching $\sim 1$ mJy. A typical
radio-loud source (with a two-point radio-optical spectral index $\alpha_{\rm
ro} \sim 0.6$) will have $V \sim 24$, beyond the reach for spectroscopy of 4m
class telescopes even in the presence of strong, broad lines, let alone if one
is dealing with a BL Lac! Similarly, at the Chandra/XMM fluxes $f_{\rm x} \sim
10^{-15}$ erg/cm$^{2}$/s a typical radio-loud source (with $\alpha_{\rm ox}
\sim 1.2$) will reach $V \sim 26$. These magnitudes are starting to become
problematic for spectral identification even for 8-10m class telescopes,
especially in the absence of strong features. I stress that these problems
will plague all radio-loud AGN and not only blazars! 

This means that we will need to be very efficient in our pre-selection of
candidates, as optical identification will require large resources.
Statistical identification of sources based on their location in
multi-parameter space, which will imply a smaller need for optical spectra
(similar to the method employed for the ``Sedentary'' survey; \S~3), will also
have to become more common. 

\section{Summary} 

The main conclusions are as follows:

\begin{enumerate}

\item ``Classical'' blazar samples are small and at relatively high fluxes; it
then follows that our understanding of the blazar phenomenon is based mostly
on the intrinsically most powerful sources. 

\item A number of on-going, deeper surveys are probing the more common, less
luminous blazars, and will reveal the bulk of the blazar population. Before
drawing conclusions about blazar demographics, however, care has to be taken
to assess the limitations of these surveys and what regions of blazar
parameter space they are sampling. 

\item Preliminary results of the Deep X-ray Radio Blazar Survey (DXRBS) in
terms of evolution, number counts, and luminosity functions agree with the
predictions of unified schemes (based on samples having flux limits $\sim 20$
times larger).

\item Even deeper blazar surveys will face daunting identification problems,
due to the faintness of the optical counterparts. The good news is, however,
that due to the relatively high radio powers of flat-spectrum quasar, we might
be approaching the limits of their Universe.

\end{enumerate}

\acknowledgements

The work on DXRBS reported here has been done in collaboration with,
amongst others, Paolo Giommi, Hermine Landt, and Eric Perlman.

\end{document}